\documentclass[onecolumn, draftclsnofoot, 12pt]{IEEEtran}
\IEEEoverridecommandlockouts
\usepackage{cite}

\usepackage[pdftex]{graphicx}
\usepackage{float}
\usepackage{fixltx2e} 

\graphicspath{{./graph/}}
\usepackage[cmex10]{amsmath}
\usepackage{epsfig,color}
\usepackage{mdwmath}
\usepackage{mdwtab}
\usepackage{amssymb}

\begin{document}
\title{On Localization of A Non-Cooperative Target with Non-Coherent Binary Detectors}
\author{Arian~Shoari,~\IEEEmembership{Student Member,~IEEE,}
        and Alireza~Seyedi,~\IEEEmembership{Senior Member,~IEEE}
       \thanks{A. Shoari is with the Department
of Electrical and Computer Engineering, University of Rochester, Rochester,
NY, E-mail: shoari@ece.rochester.edu. A. Seyedi is with the Department of Electrical Engineering and Computer Science, University of Central Florida, Orlando, FL. E-mail: alireza.seyedi@ieee.org.}}



\maketitle

\begin{abstract}
Localization of a non-cooperative target with binary detectors is considered. A general expression for the Fisher information for estimation of target location and power is developed. This general expression is then used to derive closed-form approximations for the Cram\'er-Rao bound for the case of non-coherent detectors. Simulations show that the approximations are quite consistent with the exact bounds.
\end{abstract}


\section{Introduction}
\label{secIntro}
Localization of a non-cooperative target using a number of sensors is usually performed through measurement of time difference of arrival (TDOA) \cite{Kung1998TDOA,Yang2005TDOA,Yang2006TDOA2}, direction of arrival (DOA) \cite{Kung1998TDOA}\cite{Oshman1999DOA,Kaplan2001DOA,Peng06AoA} or received signal strength (RSS) \cite{Amundson09,sheng2003collaborative,Sheng05,Blatt06,kieffer2006centralized,rabbat2012convergence,Rabbat2011,Vaghefi11_RSSbased,Rabbat10,Ampeliotis08,ArianSpawc2010,Liu2007Target,Murthy2011multiple,Murthy2012multiple,Bulusu02,VarshneyBinCoh,VarshneyFading07}. The first two approaches require complex receivers and, therefore, do not lend themselves to power and size limitations in wireless sensor networks \cite{sheng2003collaborative,Liu2007Target}. The third approach is considerably less complex and less demanding of energy, though it has a more modest performance. A good body of literature exists concerning localization of a target using RSS measurements assuming acoustic \cite{sheng2003collaborative,Sheng05,Blatt06} or radio frequency (RF) \cite{kieffer2006centralized,Vaghefi11_RSSbased,Rabbat10,rabbat2012convergence,Rabbat2011} propagation models, or without relying on any particular propagation model \cite{Ampeliotis08}.

These works assume an unquantized measurement of the received signal, or its power, in noise. In practice, however, measurements are quantized, or even binary when resources are scarce. Model independent localization schemes using noise-free binary measurements have been studied \cite{ArianSpawc2010,Liu2007Target,Murthy2011multiple,Murthy2012multiple}. In \cite{Bulusu02} a simple averaging approach for cooperative RF localization is proposed. In general, model independent approaches do not have a very good performance since they do not use the information regarding the propagation of the signal. In \cite{VarshneyBinCoh} a maximum likelihood estimator is proposed for localization of an RF target using binary signal measurements. The sensors perform coherent detection which requires phase synchronization with the target and, hence, it is not suitable for simple detectors. None of the above mentioned papers consider localization of an RF source with non-coherent detectors, which are preferred due to simplicity and low power consumption. In \cite{VarshneyFading07}, the case of non-coherent quantized observations are considered under Rayleigh fading. To the best of our knowledge, localization of a non-cooperative target using non-coherent detectors has not been studied in the absence of fading.
This scenario, considered in this paper, is important and relevant to many applications, since it models the cases where the system operates in an environment with little or no scattering, such as an open field. We first calculate the Fisher information and the Cram\'er-Rao bounds (CRB) for the estimation of location and power of the target for general binary detectors, thereby generalizing the work in \cite{ArianISIT2011} to the case where the transmit power is unknown. We then apply these results to the case of non-coherent binary detectors.

\section{Problem formulation}
\label{secProbFormulation}
Consider a non-cooperative target, located at $[x_T\,\,y_T]$, which isotropically emits energy in a two dimensional space. A large number of sensors, capable of non-coherent binary power detection, are uniformly scattered in a infinite region with density $\rho$. Each sensor makes a binary decision by comparing its received power to a threshold, $\tau$, and reports its decision, $d_i$, and its location, $[x_i ~ y_i]$, to a fusion center, where $i$ is the sensor index. The fusion center's task is to estimate the location and power of the target.

\section{Fisher Information for Localization with Binary Sensors}
\label{secFisherCRBCal}

In this section, we develop the Fisher information matrix for the estimation of location and power of the target. In other words, we would like to estimate $\boldsymbol{\theta}=[P\,\,x_T\,\,y_T]^T$, where $P$ is the power at unit distance from the target.  If $P_{\text{D},i}$ and $P_{\text{ND},i}$ denote the probabilities of detection and not detection by the $i$th sensor, respectively, the log-likelihood function can be formulated as
\begin{eqnarray}
\ln p(\{d_i\}; \boldsymbol{\theta})=\sum_{i} \left[ \delta_{i\in\mathcal{S}_{\text{D}}} \ln P_{\text{D},i} +\delta_{i\in\mathcal{S}_{\text{ND}}} \ln P_{\text{ND},i} \right],
\end{eqnarray}
where $\mathcal{S}_{\text{D}}$ ($\mathcal{S}_{\text{ND}}$) is the index set of all sensors which detect (not detect) the target, and $\delta_X$ is the indicator function of $X$. Here, we have assumed that the decision of the sensors are independent. Consequently, the Fisher information matrix given a particular set of locations $\{[x_i~y_i]\}$ is
\begin{eqnarray}
\mathbf{F}_{\{[x_i~y_i]\}}&=& -E_{\{d_i\}}\left[ \frac {\partial^2  \ln p(\{d_i\}; \boldsymbol{\theta})}{\partial \boldsymbol{\theta}^2} \right] \nonumber \\
&=& -\sum_{i} \left(E\left[\delta_{i\in\mathcal{S}_{\text{D}}} \right] \frac {\partial^2 \ln P_{\text{D},i}}{\partial \boldsymbol{\theta}^2} + E\left[\delta_{i\in\mathcal{S}_{\text{ND}}}\right] \frac {\partial^2 \ln P_{\text{ND},i}}{\partial \boldsymbol{\theta}^2}\right) \nonumber \\
&=& -\sum_{i} \left(P_{\text{D},i} \frac {\partial^2 \ln P_{\text{D},i}}{\partial \boldsymbol{\theta}^2}+ P_{\text{ND},i} \frac {\partial^2 \ln P_{\text{ND},i}}{\partial \boldsymbol{\theta}^2}\right).\nonumber
\end{eqnarray}
Thus, the contribution of the $i$th sensor to $\mathbf{F}_{\{[x_i~y_i]\}}$ is
\begin{eqnarray}
\label{IiversusIDandIND}
\mathbf{F}_i=-E_{d_i}\left[\frac {\partial^2 \ln p(d_i;\boldsymbol{\theta})}{\partial \boldsymbol{\theta}^2}\right] .
\end{eqnarray}
Expansion of the derivatives in \eqref{IiversusIDandIND} and change of coordinates to polar, with the target as origin, yields
\begin{eqnarray}
\label{Fi}
F_{i,11} &=& \frac{1}{P_{\text{D},i}(1-P_{\text{D},i})} \left(\frac {\partial {P_{\text{D},i}}}{\partial P} \right)^2  \\
F_{i,22} &=& \frac{1}{P_{\text{D},i}(1-P_{\text{D},i})} \left(\frac {\partial {P_{\text{D},i}}}{\partial r_i} \right)^2  \cos^2 \psi_i  \nonumber \\
F_{i,33} &=& \frac{1}{P_{\text{D},i}(1-P_{\text{D},i})} \left(\frac {\partial {P_{\text{D},i}}}{\partial r_i} \right)^2  \sin^2 \psi_i  \nonumber \\
F_{i,12} &=& F_{i,21}= \frac {1}{P_{\text{D},i}(1-P_{\text{D},i})} \frac {\partial {P_{\text{D},i}}}{\partial r_i} \frac {\partial {P_{\text{D},i}}}{\partial P} \cos \psi_i  \nonumber \\
F_{i,13} &=& F_{i,31}= \frac {1}{P_{\text{D},i}(1-P_{\text{D},i})} \frac {\partial {P_{\text{D},i}}}{\partial r_i} \frac {\partial {P_{\text{D},i}}}{\partial P} \sin \psi_i   \nonumber\\
F_{i,23} &=& F_{i,32}= \frac{1}{P_{\text{D},i}(1-P_{\text{D},i})} \left(\frac {\partial {P_{\text{D},i}}}{\partial r_i} \right)^2   \sin \psi_i \cos \psi_i, \nonumber
\end{eqnarray}
where $r_i$ and $\psi_i$ are the polar coordinates of the $i$th sensor relative to the target, and we have used $P_{\text{D},i}(r_i,\psi_i)=P_{\text{D},i}(r_i)$, since the propagation is assumed to be isotropic. The derivation of the above terms are straight forward but lengthy. They can be found in Appendix A.

Therefore the expected Fisher information matrix, with respect to the location of the sensors is
\begin{eqnarray}
\mathbf{F}&=&-E_{\{[x_i~y_i]\}}E_{\{d_i\}}\left[ \frac {\partial^2  \ln p(\{d_i\}; \boldsymbol{\theta})}{\partial \boldsymbol{\theta}^2} \right] \nonumber \\
&=&-E_{\{[x_i~y_i]\}} \left[ \sum_{i} P_{\text{D},i} \frac {\partial^2 \ln P_{\text{D},i}}{\partial \boldsymbol{\theta}^2}+ P_{\text{ND},i} \frac {\partial^2 \ln P_{\text{ND},i}}{\partial \boldsymbol{\theta}^2}   \right]. \nonumber
\end{eqnarray}
Calculating this expectation over $\mathbb{R}^2$ we get (see Appendix B for detail) 
\begin{eqnarray}
\mathbf{F}&=&- \int_{-\infty}^{\infty} \int_{-\infty}^{\infty} \rho \left( P_{\text{D}} \frac {\partial^2 \ln P_{\text{D}}}{\partial \boldsymbol{\theta}^2}+ P_{\text{ND}} \frac {\partial^2 \ln P_{\text{ND}}}{\partial \boldsymbol{\theta}^2} \right) dx dy  \nonumber\\
&=& -\int_{0}^{\infty} \int_0^{2 \pi}  2\rho \pi r   \left( P_{\text{D}} \frac {\partial^2 \ln P_{\text{D}}}{\partial \boldsymbol{\theta}^2}+ P_{\text{ND}} \frac {\partial^2 \ln P_{\text{ND}}}{\partial \boldsymbol{\theta}^2} \right)  dr d\phi   \nonumber
\end{eqnarray}
Now, employing (\ref{Fi}) yields
\begin{eqnarray}
\label{general_I_11}
F_{11}&=& \int_{0}^{\infty} \int_0^{2 \pi} 2 \pi \rho r \frac{\left(\frac {\partial {P_{\text{D}}}}{\partial P} \right)^2}{P_{\text{D}}(1-P_{\text{D}})} dr d\phi  \nonumber \\
&=& \int_{0}^{\infty} 4 \pi^2 \rho r  \frac{\left(\frac {\partial {P_{\text{D}}}}{\partial P} \right)^2}{P_{\text{D}}(1-P_{\text{D}})} dr,  \\
\label{general_I_22}
F_{22}=F_{33}&=& \int_{0}^{\infty} \int_0^{2 \pi} 2 \pi \rho r \frac {(\frac {\partial P_{\text{D}}}{\partial r})^2}{P_{\text{D}} (1-P_{\text{D}})} \cos^2\psi d \psi dr         \nonumber \\
&=&2\pi^2 \rho \int_{0}^{\infty}  \frac {(\frac {\partial P_{\text{D}}}{\partial r})^2}{P_{\text{D}} (1-P_{\text{D}})} rdr.
\end{eqnarray}
And for the off-diagonal terms in $\mathbf{F}$ we have,
\begin{eqnarray}
\label{generalE_I_23}
F_{12} &=& \int_{0}^{\infty} \int_0^{2 \pi} 2 \pi \rho r \frac{\frac {\partial {P_{\text{D}}}}{\partial r}  \frac {\partial {P_{\text{D}}}}{\partial P} }{P_{\text{D}}(1-P_{\text{D}})}  \cos \psi  d \psi dr         \nonumber \\
&=&\int_{0}^{\infty}  2 \pi \rho r \frac{\frac {\partial {P_{\text{D}}}}{\partial r} \frac {\partial {P_{\text{D}}}}{\partial P} }{P_{\text{D}}(1-P_{\text{D}})}  \left(\int_0^{2 \pi} \cos \psi  d \psi\right)  dr = 0. \nonumber
\end{eqnarray}
Similarly, $F_{21}=F_{13}=F_{31}=F_{32}=F_{23}=0.$
In other words, the Fisher information matrix is diagonal and the corresponding CRBs are
\begin{eqnarray}
C_{P}=\frac{1}{F_{11}}, C_{x}=C_{y}=\frac{1}{F_{22}}=\frac{1}{F_{33}}.  \nonumber
\end{eqnarray}
Note that when $P$ is known the Fisher information matrix will be a sub-matrix of $\mathbf{F}$ containing only the second and the third rows and columns, which is also diagonal. Therefore the $C_{x}$ and $C_{y}$ are the same as that when $P$ is unknown. In other words (for binary sensors) the knowledge of $P$ does not affect the CRBs for $x_T$ and $y_T$. Note that this result is independent of the propagation model and the type of detection.

\section{Fisher Information and CRB for Localization with Non-Coherent Binary Sensors}
\label{secProbDetection}
\subsection{Probability of Detection}
\vspace{-0.05in}
Assuming a narrowband transmission, the received signal at the $i$th detector can be modeled by
\begin{eqnarray}
s_i(t)&=&\sqrt{\frac {2P}{r_i^{\alpha}}} \sin(\omega t+\phi_i)+n_i(t), \nonumber
\end{eqnarray}
where $\omega$ is the frequency, $r_i$ is the distance of the $i$th sensor from the target, $\alpha$ is the pathloss exponent, and $n_i(t)$ is a white Gaussian noise process with power $\sigma^2$. We assume the phase, $\phi_i$, is uniformly distributied over $[0,2\pi)$. The amplitude and phase of the received signal for each sensor are unknown. With this model, the optimal detection rule is
\begin{eqnarray}
U_i & \underset{0}{\overset{1}{\gtrless}} & \frac{2\tau}{\sigma^2},\nonumber
\end{eqnarray}
where $U_i$ is a non-central chi-squared variable with non-centrality parameter $\lambda_i=\frac {TP}{\sigma^2r_i^{ \alpha}}$ and two degrees of freedom, and $T$ is the duration of the measurement (see Sec. 7.6.2 of \cite{Kay93Detectionbook}). Thus, the probability of detection for the $i$th sensor is
\begin{eqnarray}
\label{PD_i}
P_{\text{D},i} &=& Pr \left\{ \frac{\sigma^2U_i}{2} > \tau \right\} = Q \left(\sqrt{\frac {T P}{\sigma^2 r_i^{\alpha}} }, \sqrt{\frac {2\tau}{\sigma^2}}\right),
\end{eqnarray}
where $Q(.,.)$ is the Marcum Q function \cite{NuttallQM1975}. Thus, the probability of decision sequence $\{d_i\}$ is
\begin{eqnarray}
\label{MLFormulaforall}
p\left(\{d_i\};\boldsymbol{\theta}\right) & =& \prod_{i \in \mathcal{S}_{\text{D}}} Q\left(\sqrt{\frac {T P}{\sigma^2 r_i^{\alpha}} },\sqrt{\frac {2\tau}{\sigma^2}}\right) \times
\prod_{i \in \mathcal{S}_{\text{ND}}} \left[1- Q\left(\sqrt{\frac {T P}{\sigma^2 r_i^{\alpha}} },\sqrt{\frac {2\tau}{\sigma^2}} \right) \right]. \nonumber
\end{eqnarray}

\subsection{Fisher Information and CRB}
\vspace{-0.05in}
To calculate the partial derivatives of $P_{\text{D},i}$ with respect to $P$ and $r_i$ we recall that the partial derivative of the Marcum Q function with respect to its first argument is given by \cite{PrattPartDerQ1968}
\begin{eqnarray}
\frac {\partial Q(a,b)}{\partial a} &=& b I_1(ab)e^{-\frac{a^2+b^2}{2}}.	\nonumber	
\end{eqnarray}
Thus,
\begin{eqnarray}
\label{rgD_rP0_eqQ1}
\frac {\partial P_{\text{D},i}}{\partial r_i}&=&-\frac{\alpha}{2r_i}\sqrt{\frac {2T P\tau}{\sigma^4r_i^{\alpha}}}I_1\left(\sqrt{\frac {2T P\tau}{\sigma^4 r_i^{\alpha}}}\right)e^{-\frac{T P}{2\sigma^2 r_i^{\alpha}} -\frac {\tau}{\sigma^2}}\nonumber\\
\frac {\partial P_{\text{D},i}}{\partial P}&=&\frac{1}{2P}\sqrt{\frac {2T P\tau}{\sigma^4 r_i^{\alpha}}}I_1\left(\sqrt{\frac {2T P\tau}{\sigma^4 r_i^{\alpha}}}\right)e^{-\frac{T P}{2\sigma^2 r_i^{\alpha}} - \frac {\tau}{\sigma^2}}.
\end{eqnarray}

The elements of fisher information matrix can be calculated by substituting (\ref{PD_i}) and (\ref{MLFormulaforall}) in (\ref{general_I_11}) and (\ref{general_I_22}). However, we note that the power law propagation model is not realistic for short distances \cite{rappaport1996wireless}. To get around this problem, we use the fact that the probability of having a sensor very close to the target is small. Hence, we start the integration in (\ref{general_I_11}) and (\ref{general_I_22}) from $\breve{r}= \frac{1}{\sqrt{4\rho}}$ which is the average distance of the closest detector to the target (see Appendix C). That is,
\begin{eqnarray}
\label{Approx_I_11}
F_{11}&\approx& \int_{\breve{r}}^{\infty} 4 \pi^2 \rho r  \frac{\left(\frac {\partial {P_{\text{D}}}}{\partial P} \right)^2}{P_{\text{D}}(1-P_{\text{D}})} dr.\\
F_{22}=F_{33}&\approx& 2\pi^2 \rho \int_{\breve{r}}^{\infty}  \frac {(\frac {\partial P_{\text{D}}}{\partial r})^2}{P_{\text{D}} (1-P_{\text{D}})} rdr,
\label{Approx_I_22}
\end{eqnarray}

Now, change of variables $x=\sqrt{\frac {T P}{\sigma^2 r^{\alpha}} }$ and $t=\sqrt{\frac {2\tau}{\sigma^2}}$, reduces (\ref{rgD_rP0_eqQ1}) to
\begin{eqnarray}
\label{rPD_rr}
\frac {\partial {P_{\text{D}}}}{\partial r}&=& \frac{-\alpha t x}{2r}  e^{-\frac {t^2+x^2}{2}} I_1(tx),\nonumber\\
\frac {\partial {P_{\text{D}}}}{\partial P}&=& \frac{tx}{2P}  e^{-\frac {t^2+x^2}{2}} I_1(tx)
\end{eqnarray}
and (\ref{Approx_I_11}) becomes
\begin{eqnarray}
\label{Eefinalaccurateintegral}
F_{11} \approx \frac{2\pi^2t^2\rho T^{\frac{2}{\alpha}}P^{\frac{2}{\alpha}-2}}{\alpha \sigma^{\frac{4}{\alpha}}} \int_{0}^{\breve{x}}     \frac{x^{1-\frac{4}{\alpha}}e^{-x^2-t^2}I_1^2(tx)}{Q(x,t)(1-Q(x,t))} dx,
\end{eqnarray}
where $\breve{x}=\sqrt{\frac {T P}{\sigma^2 {\breve{r}}^{\alpha}} }$. Unfortunately (\ref{Eefinalaccurateintegral}) is intractable. Hence, in the following we provide an approximation. Let us define
\begin{eqnarray}
\label{fxt}
f(x,t) =  \ln \frac{1}{Q(x,t)[1-Q(x,t)]}  \approx  f_0(t)+f_1(t)x+f_2(t)x^2, \nonumber
\end{eqnarray}
where the approximation is performed by neglecting the tail of the Taylor expansion with respect to $x$ around $x=\breve{x}$, and
\begin{eqnarray}
\label{f0f1f2}
f_0(t) &=& f(\breve{x},t)-\breve{x} f'(\breve{x},t)+\frac {\breve{x}^2}{2} f''(\breve{x},t) \nonumber \\
f_1(t) &=& f'(\breve{x},t)-\breve{x} f''(\breve{x},t) \nonumber \\
f_2(t) &=& \frac{1}{2} f''(\breve{x},t),  \nonumber
\end{eqnarray}
where
\begin{eqnarray}
\label{fpr_fzg}
f'(\breve{x},t)&=& -\frac{Q'(\breve{x},t)}{Q(\breve{x},t)}+\frac{Q'(\breve{x},t)}{1-Q(\breve{x},t)}         \nonumber \\
f''(\breve{x},t)&=&\frac{Q'(\breve{x},t)^2-Q(\breve{x},t) Q''(\breve{x},t)}{Q(\breve{x},t)^2}+\frac{Q'(\breve{x},t)^2+(1-Q(\breve{x},t))Q''(\breve{x},t)) }{(1-Q(\breve{x},t))^2}  \nonumber\\
Q'(\breve{x},t)&=& tI_1(\breve{x} t)e^{-\frac{\breve{x}^2+t^2}{2}}    \nonumber \\
Q''(\breve{x},t)&=& \left[ \frac{t^2}{2}I_0(\breve{x}t)-t \breve{x}I_1(t \breve{x})+ \frac{t^2}{2} I_2(t \breve{x})  \right] e^{-\frac{\breve{x}^2+t^2}{2}}. \nonumber
\end{eqnarray}
Therefore, the integrand of (\ref{Eefinalaccurateintegral}) is
\begin{eqnarray}
\label{Jmainformula}
h(x,t) & = & \frac{x^{1-\frac{4}{\alpha}}e^{-x^2-t^2}I_1^2(xt)}{Q(x,t)[1-Q(x,t)]}\nonumber\\
&\approx& x^{1-\frac{4}{\alpha}} I_1^2(xt)e^{-x^2-t^2}e^{f_0(t)+f_1(t)x+f_2(t)x^2}. \nonumber
\end{eqnarray}
Change of variables $y=xt$ yields
\begin{eqnarray}
h(y,t) &=& t^{-1+\frac{4}{\alpha}}y^{1-\frac{4}{\alpha}} I_1^2(y) e^{(f_0(t)-t^2)+\frac{f_1(t)}{t}y+\frac{(f_2(t)-1)}{t^2} y^2} \nonumber \\
&=& C t^{-1+\frac{4}{\alpha}}y^{1-\frac{4}{\alpha}}  I_1^2(y) e^{-B^2-2AB y- A^2y^2},\nonumber
\end{eqnarray}
where $A= \frac{\sqrt{1-f_2(t)}}{t}$, $B= -\frac{f_1(t)}{2 \sqrt {1-f_2(t)}}$ and  $C= \exp\left(\frac{f_1(t)^2}{4 |1-f_2(t)| } + f_0(t)-t^2\right)$. Thus,
\begin{eqnarray}
\label{J1approximation}
F_{11} &\approx& \frac{2C\pi^2\rho T^{\frac{2}{\alpha}}t^{\frac{4}{\alpha}}P^{\frac{2}{\alpha}-2}}{\alpha \sigma^{\frac{4}{\alpha}}}   \int_{0}^{\breve{y}}     y^{1-\frac{4}{\alpha}}  I_1^2(y) e^{-(Ay+B)^2} dy,\nonumber
\end{eqnarray}
where $\breve{y}=\breve{x}t$. We also approximate $I_1^2(y)$ with the first $m+1$ terms of its Taylor expansion around zero \cite{bender2003powers}
\begin{eqnarray}
\label{TaylorExpansionI1sq}
I_1^2(y) & \approx & \sum_{k=0}^{m} {2k+2\choose k} \left[\frac{y^{k+1}}{2^{k+1}(k+1)!}\right]^2. \nonumber
\end{eqnarray}
Therefore,
\begin{eqnarray}
\label{Ee2integralapprox}
F_{11} &\approx& \frac{2C\pi^2\rho T^{\frac{2}{\alpha}}t^{\frac{4}{\alpha}}P^{\frac{2}{\alpha}-2}}{\alpha \sigma^{\frac{4}{\alpha}}}  \sum_{k=0}^{m} {2k+2\choose k} \times \frac{1}{\left[2^{k+1}(k+1)!\right]^2}\int_0^{\breve{y}}     y^{2k+3-\frac{4}{\alpha}}   e^{-(Ay+B)^2}  dy \nonumber\\
\label{Ee2integralapprox12}
&=& \frac{2C\pi^2\rho T^{\frac{2}{\alpha}}t^{\frac{4}{\alpha}}P^{\frac{2}{\alpha}-2}}{\alpha A^{2-\frac{4}{\alpha}} \sigma^{\frac{4}{\alpha}}}   \sum_{k=0}^{m} {2k+2\choose k}\times\frac{1}{\left[(2A)^{k+1}(k+1)!\right]^2}\int_B^{\breve{s}}    \left(s-B\right)^{2k+3-\frac{4}{\alpha}}   e^{-s^2}  ds,\nonumber
\end{eqnarray}
where $s=Ay+B$ and $\breve{s}=A\breve{y}+B$. If $\alpha = 2$ or $4$, the integral above is the scaled partial moment of a Gaussian with respect to $B$ and can be represented in terms of incomplete gamma functions. For $\alpha=2$ we get
\begin{eqnarray}
F_{11} &\approx& \frac{C\pi^2\rho Tt^2}{2 P\sigma^2}   \sum_{k=0}^{m} {2k+2\choose k} \frac{1}{\left[(2A)^{k+1}(k+1)!\right]^2}   \nonumber\\
&  & \times \sum_{l=0}^{2k+1}{2k+1\choose l}(-B)^{l} \left[ \Gamma\left(k+1-\frac{l}{2},B^2\right)\right.\left.-\Gamma\left(k+1-\frac{l}{2},\breve{s}^2\right) \right],\nonumber
\end{eqnarray}
where $\Gamma(.,.)$ is the upper incomplete gamma function. Similarly, for $\alpha=4$ we have
\begin{eqnarray}
F_{11} &\approx& \frac{C\pi^2\rho \sqrt{T} t}{4 A P^{\frac{3}{2}} \sigma}   \sum_{k=0}^{m} {2k+2\choose k} \frac{1}{\left[(2A)^{k+1}(k+1)!\right]^2}  \nonumber\\
& & \times  \sum_{l=0}^{2k+2}{2k+2\choose l}(-B)^{l} \left[ \Gamma\left(k+1+\frac{1-l}{2},B^2\right) - \Gamma\left(k+1+\frac{1-l}{2},\breve{s}^2\right) \right]. \nonumber
\end{eqnarray}

We can calculate $F_{22}$ and $F_{33}$ by substituting (\ref{PD_i}) and (\ref{rgD_rP0_eqQ1}) into (\ref{Approx_I_22}) and change of variables $x=\sqrt{\frac {T P}{\sigma^2 r^{\alpha}} }$ and $t=\sqrt{\frac {2\tau}{\sigma^2}}$
\begin{eqnarray}
\label{NonCoh_I_22_xt}
F_{22}=F_{33}\approx\pi^2 \rho \alpha t^2 \int_0^{\breve{x}}   \frac {I_1^2\left(xt\right)e^{-x^2-t^2}}{Q(x,t) (1-Q(x,t))} xdx
\end{eqnarray}
Note that $P$ only appears in $\breve{x}=\sqrt{\frac {T P}{\sigma^2 {\breve{r}}^{\alpha}} }$. As the sensor density $\rho$ increases, $\check{r}$ vanishes, and $F_{22}$ becomes independent of $P$. Note that this does not mean that the performance of a location estimator is
independent of $P$, since the estimator may not be efficient.
Derivations similar to those used for $F_{11}$ reduce (\ref{NonCoh_I_22_xt}) to (see Appendix D for detail) 
\begin{eqnarray}
\label{I22Generalexac2}
F_{22}&\approx&\frac{C\pi^2 \rho \alpha }{2A^2} \sum_{k=0}^m {2k+2\choose k}\frac{1}{[(2A)^{k+1}(k+1)!]^2}    \nonumber\\
& & \times \sum_{l=0}^{2k+3}{2k+3\choose l}(-B)^{l} \left[ \Gamma\left(k+2-\frac{l}{2},B^2\right)- \Gamma\left(k+2-\frac{l}{2},{\breve{s}}^2\right) \right].\nonumber
\end{eqnarray}

\section{Numerical Results}
\label{secSimulation}
\begin{figure}[!t]
\begin{centering}
\includegraphics[width=4in]{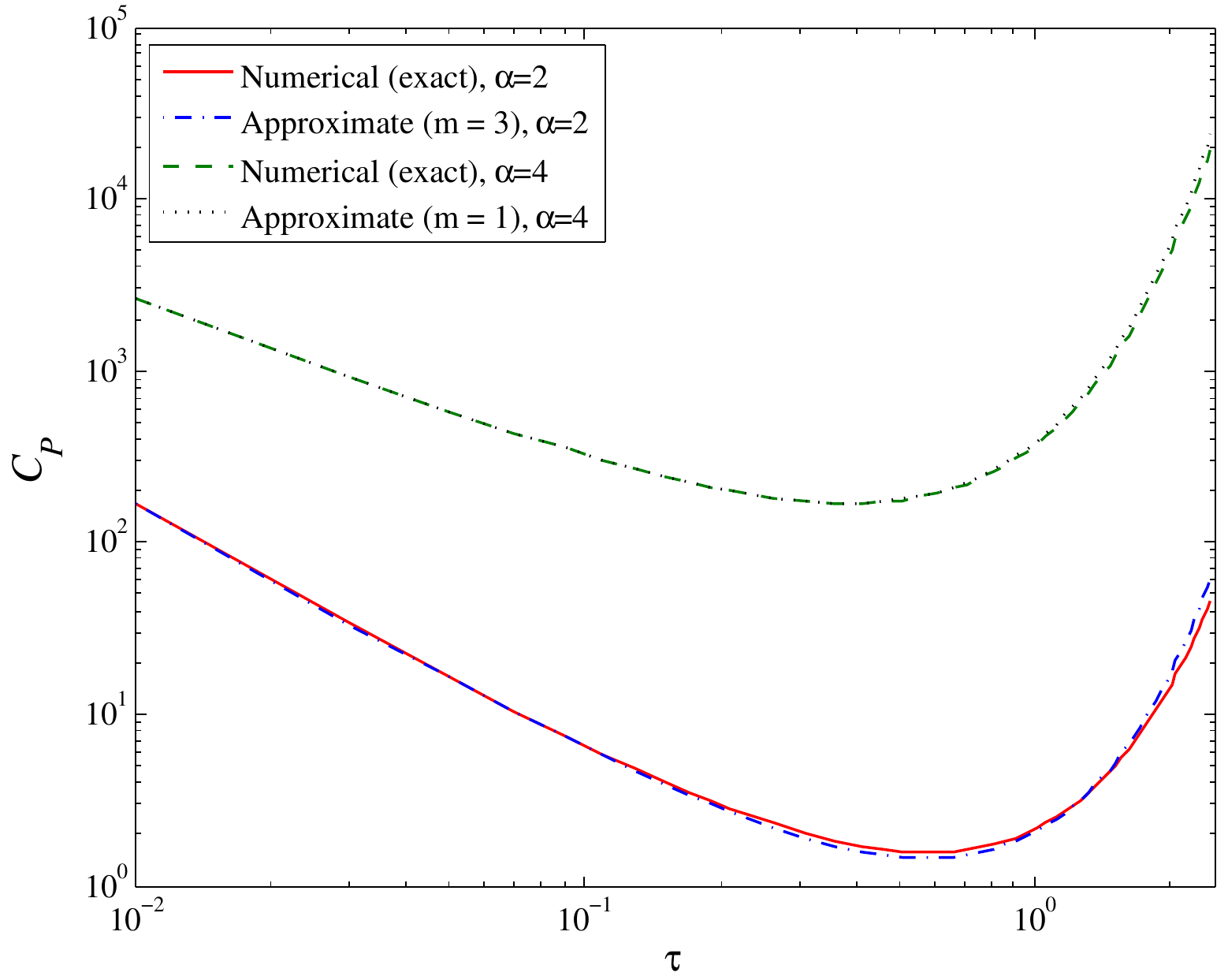}
\caption{CRB for estimation of $P$ versus $\tau$.}
\label{CRB_11_Paper_1}
\end{centering}
\end{figure}
\begin{figure}[!t]
\begin{centering}
\includegraphics[width=4in]{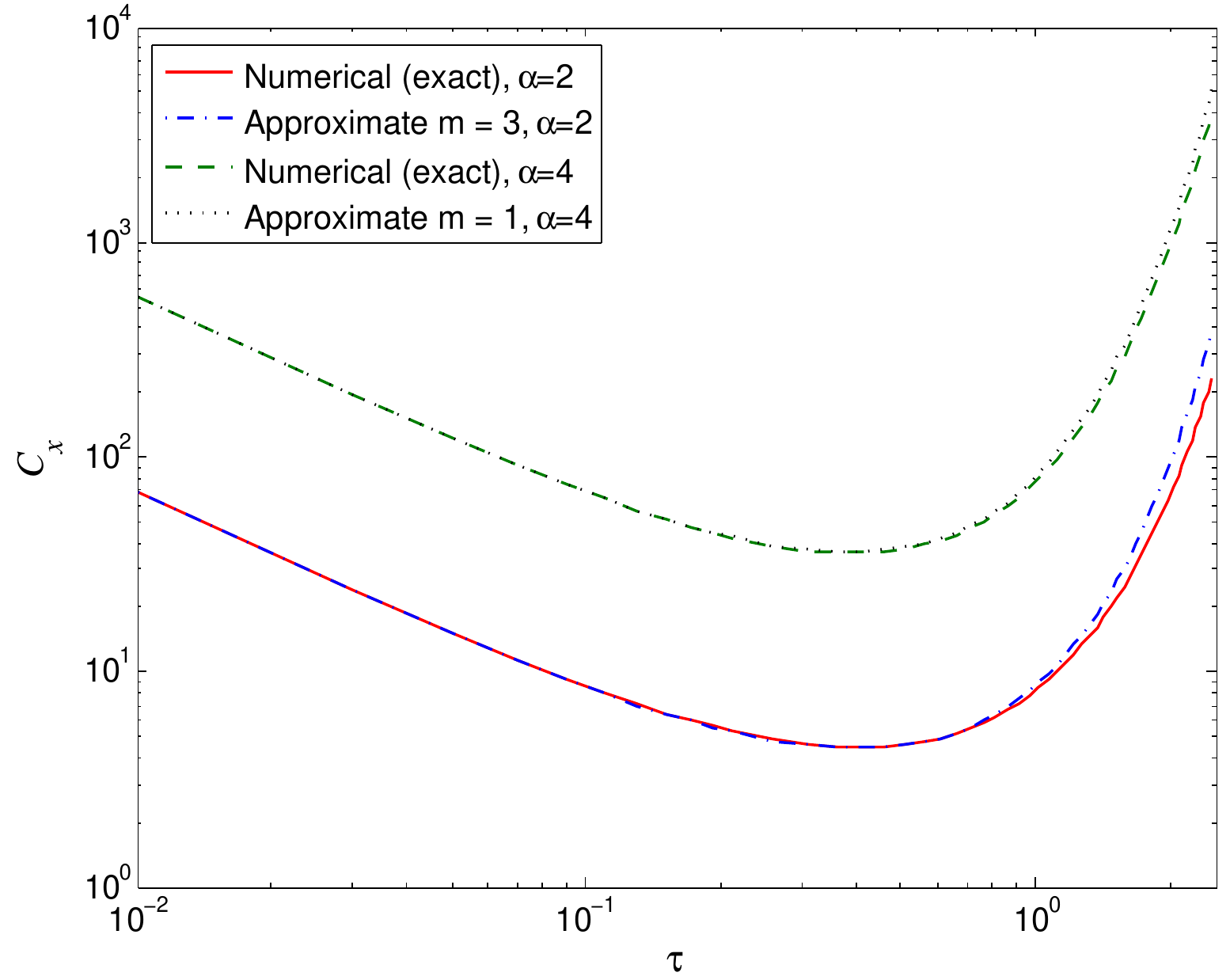}
\caption{CRB for estimation of $x_T$ versus $\tau$.}
\label{CRB_22_Paper_2}
\end{centering}
\end{figure}
In this section, we compare the developed closed-form approximations of the CRBs with their exact numerical calculation. We consider $P=2$, $T=1$, $\sigma=0.5$ and $\rho=0.05$. Fig. \ref{CRB_11_Paper_1} depicts $C_{P}$ for pathloss exponents $\alpha=2$ and $4$. Similarly, Fig. \ref{CRB_22_Paper_2} presents $C_{x}$ for pathloss exponents $\alpha=2$ and $4$. We note that, again, an optimum threshold exists, which is only slightly different from that of the estimation of $P$. We also see that $m =1$ and $m=3$ are sufficient for excellent approximations for $\alpha=4$ and $\alpha=2$, respectively. In both figures we see that there exists an optimum value of $\tau$ that provides the best performance and the performance deteriorates at smaller and larger values of $\tau$. When $\tau$ is too small, many sensors incorrectly detect the target (analogous to \emph{false alarm}). Thus the poor quality of local decisions leads to poor overall performance. When $\tau$ is too large, the decisions of the sensors are more accurate, but only few sensors will detect the target. Consequently, it is difficult to estimate the power or the location of the target using these few detecting sensors.

\section{Conclusion}
The problem of localization of an uncooperative target with non-coherent binary detectors in the absence of fading is studied.
Fisher information matrix and CRBs for the localization with binary detectors derived. This general result is then employed to calculate the  Fisher information and CRBs for the case of non-coherent detectors. It is shown that there is an optimum detection threshold value which correspond to the lowest CRBs. We also note that for large node densities, CRB of location estimation becomes independent of the transmitter power. Future research directions include extention to allow different local threshold for the sensors, and derivation of a closed form expression for the optimal threshold.

\appendices
\section{Derivation of (\ref{Fi})} 
We have
\begin{eqnarray}
\mathbf{F}_{\{[x_i~y_i]\}}&=& -E_{\{d_i\}}\left[ \frac {\partial^2  \ln p(\{d_i\}; \boldsymbol{\theta})}{\partial \boldsymbol{\theta}^2} \right] \nonumber \\
&=& -\sum_{i} P_{\text{D},i} \frac {\partial^2 \ln P_{\text{D},i}}{\partial \boldsymbol{\theta}^2}+ P_{\text{ND},i} \frac {\partial^2 \ln P_{\text{ND},i}}{\partial \boldsymbol{\theta}^2}. \nonumber
\end{eqnarray}
Noting that $P_{\text{ND},i}=1- P_{\text{D},i}$, the contribution of the $i$th sensor to $\mathbf{F}_{\{[x_i~y_i]\}}$ is
\begin{eqnarray}
\label{IiversusIDandIND}
\mathbf{F}_i&=&-E_{d_i}\left[\frac {\partial^2 \ln p(d_i;\boldsymbol{\theta})}{\partial \boldsymbol{\theta}^2}\right]   \nonumber \\
&=&  \left[ \begin{array}{ccc} F_{i,11} & F_{i,12} & F_{i,13} \\
F_{i,21} & F_{i,22} & F_{i,23} \\
F_{i,31} & F_{i,32} & F_{i,33} \end{array}\right] \nonumber \\
&=& P_{\text{D},i} \mathbf{F}_{\text{D},i} + (1-P_{\text{D},i}) \mathbf{F}_{\text{ND},i},\nonumber
\end{eqnarray}
where the elements of $\mathbf{F}_{\text{D},i}=[F_{\text{D},i,mn}]$ and $\mathbf{F}_{\text{ND},i}=[F_{\text{ND},i,mn}]$ are

\begin{eqnarray}
\label{element1expansion}
F_{\text{D},i,11} =   -\frac{\partial^2 \ln P_{\text{D},i}}{\partial P^2} &=& -\frac{\partial}{\partial P} \left( \frac{1}{P_{\text{D},i}} \frac{\partial P_{\text{D},i}}{\partial P}  \right) \nonumber \\
&=& \frac{1}{P_{\text{D},i}^2} \left(\frac {\partial {P_{\text{D},i}}}{\partial P} \right)^2   -   \frac{1}{P_{\text{D},i}} \frac {\partial^2 {P_{\text{D},i}}}{\partial P^2},  \nonumber \\
F_{\text{D},i,22} =  -\frac{\partial^2 \ln P_{\text{D},i}}{\partial x_T^2}&=& -\frac{\partial}{\partial x_T} \left( \frac{1}{P_{\text{D},i}} \frac{\partial P_{\text{D},i}}{\partial x_T}  \right)        \nonumber \\
&=&\frac{1}{P_{\text{D},i}^2} \left(\frac {\partial {P_{\text{D},i}}}{\partial x_T} \right)^2   -   \frac{1}{P_{\text{D},i}} \frac {\partial^2 {P_{\text{D},i}}}{\partial x_T^2}, \nonumber \\
F_{\text{D},i,33} =  -\frac{\partial^2 \ln P_{\text{D},i}}{\partial y_T^2}&=& -\frac{\partial }{\partial y_T} \left(\frac{1}{P_{\text{D},i}} \frac{\partial P_{\text{D},i}}{\partial y_T} \right) \nonumber \\
&=&\frac{1}{P_{\text{D},i}^2} \left(\frac {\partial {P_{\text{D},i}}}{\partial y_T} \right)^2   -   \frac{1}{P_{\text{D},i}} \frac {\partial^2 {P_{\text{D},i}}}{\partial y_T^2},        \nonumber \\
F_{\text{D},i,12} = F_{\text{D},i,21}= -\frac{\partial^2 \ln P_{\text{D},i}}{\partial x_T \partial P}&=& -\frac{\partial}{\partial x_T} \left( \frac{1}{P_{\text{D},i}} \frac{\partial P_{\text{D},i}}{\partial P} \right)   \nonumber \\
&=&\frac{1}{P_{\text{D},i}^2} \frac {\partial {P_{\text{D},i}}}{\partial x_T} \frac {\partial {P_{\text{D},i}}}{\partial P}   -   \frac{1}{P_{\text{D},i}} \frac {\partial^2 {P_{\text{D},i}}}{\partial x_T \partial P}, \nonumber \\
F_{\text{D},i,13} = F_{\text{D},i,31}= -\frac{\partial^2 \ln P_{\text{D},i}}{\partial y_T \partial P}&=& -\frac{\partial}{\partial y_T} \left(\frac{1}{P_{\text{D},i}} \frac{\partial P_{ \text{D},i}}{\partial P} \right) \nonumber \\
&=&\frac{1}{P_{\text{D},i}^2} \frac {\partial {P_{\text{D},i}}}{\partial y_T} \frac {\partial {P_{\text{D},i}}}{\partial P}   -   \frac{1}{P_{\text{D},i}} \frac {\partial^2 {P_{\text{D},i}}}{\partial y_T \partial P},    \nonumber \\
F_{\text{D},i,23} = F_{\text{D},i,32}=-\frac{\partial^2 \ln P_{\text{D},i}}{\partial x_T \partial y_T}&=& -\frac{\partial}{\partial x_T} \left(\frac{1}{P_{\text{D},i}} \frac{\partial P_{\text{D},i}}{\partial y_T} \right) \nonumber \\
&=& \frac{1}{P_{\text{D},i}^2} \frac {\partial {P_{\text{D},i}}}{\partial x_T} \frac {\partial {P_{\text{D},i}}}{\partial y_T}   -   \frac{1}{P_{\text{D},i}} \frac {\partial^2 {P_{\text{D},i}}}{\partial x_T \partial y_T},\nonumber
\end{eqnarray}
and
\begin{eqnarray}
\label{element2expansion}
F_{\text{ND},i,11} =  -\frac{\partial^2 \ln P_{\text{ND},i}}{\partial P^2}&=& -\frac{\partial^2 \ln (1-P_{\text{D},i})}{\partial P^2} =-\frac{\partial}{\partial P} \left( \frac{1}{1-P_{\text{D},i}} \frac{-\partial P_{\text{D},i}}{\partial P} \right)  \nonumber \\
&=& \frac{1}{(1-P_{\text{D},i})^2} \left(\frac {\partial {P_{\text{D},i}}}{\partial P} \right)^2   +   \frac{1}{1-P_{\text{D},i}} \frac {\partial^2 {P_{\text{D},i}}}{\partial P^2},   \nonumber \\
F_{\text{ND},i,22} = -\frac{\partial^2 \ln P_{\text{ND},i}}{\partial x_T^2}&=&-\frac{\partial^2 \ln (1-P_{\text{D},i})}{\partial x_T^2}  = -\frac{\partial}{\partial x_T} \left(\frac{1}{1-P_{\text{D},i}} \frac{-\partial P_{\text{D},i}}{\partial x_T} \right)  \nonumber \\
&=&\frac{1}{(1-P_{\text{D},i})^2} \left(\frac {\partial {P_{\text{D},i}}}{\partial x_T} \right)^2   +   \frac{1}{1-P_{\text{D},i}} \frac {\partial^2 {P_{\text{D},i}}}{\partial x_T^2}, \nonumber \\
F_{\text{ND},i,33} =  -\frac{\partial^2 \ln P_{\text{ND},i}}{\partial y_T^2}&=& -\frac{\partial^2 \ln (1-P_{\text{D},i})}{\partial y_T^2}=-\frac{\partial}{\partial y_T} \left(\frac{1}{1-P_{\text{D},i}} \frac{-\partial P_{\text{D},i}}{\partial y_T}   \right) \nonumber
\end{eqnarray}

\begin{eqnarray}
&=&\frac{1}{(1-P_{\text{D},i})^2} \left(\frac {\partial {P_{\text{D},i}}}{\partial y_T} \right)^2   +   \frac{1}{1-P_{\text{D},i}} \frac {\partial^2 {P_{\text{D},i}}}{\partial y_T^2},  \nonumber \\
F_{\text{ND},i,12}=F_{\text{ND},i,21}=-\frac{\partial^2 \ln P_{\text{ND},i}}{\partial x_T \partial P}&=&-\frac{\partial^2 \ln (1-P_{\text{D},i})}{\partial x_T \partial P}=-\frac{\partial}{\partial x_T} \left(\frac{1}{1-P_{\text{D},i}} \frac{-P_{\text{D},i}}{\partial P} \right)  \nonumber \\
&=&\frac{1}{(1-P_{\text{D},i})^2}  \frac {\partial {P_{\text{D},i}}}{\partial x_T}  \frac {\partial {P_{\text{D},i}}}{\partial P}   +   \frac{1}{1-P_{\text{D},i}} \frac {\partial^2 {P_{\text{D},i}}}{\partial x_T \partial P}, \nonumber \\
F_{\text{ND},i,13}=F_{\text{ND},i,31}=-\frac{\partial^2 \ln P_{\text{ND},i}}{\partial y_T \partial P} &=& -\frac{\partial^2 \ln (1-P_{\text{D},i})}{\partial y_T \partial P} =\frac{\partial}{\partial y_T} \left(\frac{1}{1-P_{\text{D},i}} \frac{-P_{\text{D},i}}{\partial P}   \right) \nonumber \\
&=& \frac{1}{(1-P_{\text{D},i})^2}  \frac {\partial {P_{\text{D},i}}}{\partial y_T}  \frac {\partial {P_{\text{D},i}}}{\partial P}   +   \frac{1}{1-P_{\text{D},i}} \frac {\partial^2 {P_{\text{D},i}}}{\partial y_T \partial P}, \nonumber \\
F_{\text{ND},i,23}=F_{\text{ND},i,32}=-\frac{\partial^2 \ln P_{\text{ND},i}}{\partial x_T \partial y_T}&=&-\frac{\partial^2 \ln (1-P_{\text{D},i})}{\partial x_T \partial y_T}=-\frac{\partial}{\partial x_T}\left(\frac{1}{1-P_{\text{D},i}} \frac{-\partial P_{\text{D},i}}{\partial y_T} \right)  \nonumber \\
&=&\frac{1}{(1-P_{\text{D},i})^2}  \frac {\partial {P_{\text{D},i}}}{\partial x_T}  \frac {\partial {P_{\text{D},i}}}{\partial y_T}   +   \frac{1}{1-P_{\text{D},i}} \frac {\partial^2 {P_{\text{D},i}}}{\partial x_T \partial y_T}.\nonumber
\end{eqnarray}
Substitution of of these elements in $\mathbf{F}_i$ yields
\begin{eqnarray}
\label{elementsexpansionall}
F_{i,11}&=& P_{\text{D},i} F_{\text{D},i,11} + (1-P_{\text{D},i}) F_{\text{ND},i,11}= \frac{1}{P_{\text{D},i}(1-P_{\text{D},i})} \left(\frac {\partial {P_{\text{D},i}}}{\partial P} \right)^2,  \nonumber \\
F_{i,22} &=& P_{\text{D},i} F_{\text{D},i,22} + (1-P_{\text{D},i}) F_{\text{ND},i,22}=\frac{1}{P_{\text{D},i}(1-P_{\text{D},i})} \left(\frac {\partial {P_{\text{D},i}}}{\partial x_T} \right)^2,  \nonumber \\
F_{i,33} &=& P_{\text{D},i} F_{\text{D},i,33} + (1-P_{\text{D},i}) F_{\text{ND},i,33}= \frac{1}{P_{\text{D},i}(1-P_{\text{D},i})} \left(\frac {\partial {P_{\text{D},i}}}{\partial y_T} \right)^2,   \nonumber \\
F_{i,12}&=&F_{i,21}= P_{\text{D},i} F_{\text{D},i,12} + (1-P_{\text{D},i}) F_{\text{ND},i,12}= \frac{1}{P_{\text{D},i}(1-P_{\text{D},i})} \frac {\partial {P_{\text{D},i}}}{\partial x_T} \frac {\partial {P_{\text{D},i}}}{\partial P},  \nonumber \\
F_{i,13}&=&F_{i,31}= P_{\text{D},i} F_{\text{D},i,13} + (1-P_{\text{D},i}) F_{\text{ND},i,13}= \frac{1}{P_{\text{D},i}(1-P_{\text{D},i})} \frac {\partial {P_{\text{D},i}}}{\partial y_T} \frac {\partial {P_{\text{D},i}}}{\partial P}, \nonumber \\
F_{i,23} &=& F_{i,32}= P_{\text{D},i} F_{\text{D},i,23} + (1-P_{\text{D},i}) F_{\text{ND},i,23}= \frac{1}{P_{\text{D},i}(1-P_{\text{D},i})} \frac {\partial {P_{\text{D},i}}}{\partial x_T} \frac {\partial {P_{\text{D},i}}}{\partial y_T}.\nonumber
\end{eqnarray}
Since we have assumed the propagation is isotropic, we have $P_{\text{D},i}(r_i,\psi_i)=P_{\text{D},i}(r_i)$ where $r_i$ and $\psi_i$ are the polar coordinates of the $i$th sensor relative to the target. This means that
\begin{eqnarray}
\frac {\partial {P_{\text{D},i}}}{\partial x_T}=\frac {\partial {P_{\text{D},i}}}{\partial r_i}\frac {\partial r_i}{\partial x_T},  \nonumber \\
\frac {\partial {P_{\text{D},i}}}{\partial y_T}=\frac {\partial {P_{\text{D},i}}}{\partial r_i}\frac {\partial r_i}{\partial y_T}.  \nonumber
\end{eqnarray}
Moreover,
\begin{eqnarray}
\frac {\partial r_i}{\partial x_T}=\frac {\partial \sqrt{(x_i-x_T)^2+(y_i-y_T)^2}}{\partial x_T}=\frac{x_i-x_T}{\sqrt{(x_i-x_T)^2+(y_i-y_T)^2}}=\cos \psi_i, \nonumber\\
\frac {\partial r_i}{\partial y_T}=\frac {\partial \sqrt{(x_i-x_T)^2+(y_i-y_T)^2}}{\partial y_T}=\frac{y_i-y_T}{\sqrt{(x_i-x_T)^2+(y_i-y_T)^2}}=\sin \psi_i.\nonumber
\end{eqnarray}
Thus,
\begin{eqnarray}
\label{elementsexpansionallpsi}
F_{i,11} &=& \frac{1}{P_{\text{D},i}(1-P_{\text{D},i})} \left(\frac {\partial {P_{\text{D},i}}}{\partial P} \right)^2, \nonumber \\
F_{i,22} &=& \frac{1}{P_{\text{D},i}(1-P_{\text{D},i})}\left(\frac {\partial {P_{\text{D},i}}}{\partial r_i} \right)^2   \cos^2 \psi_i,  \nonumber \\
F_{i,33} &=& \frac{1}{P_{\text{D},i}(1-P_{\text{D},i})} \left(\frac {\partial {P_{\text{D},i}}}{\partial r_i} \right)^2  \sin^2 \psi_i,  \nonumber \\
F_{i,12} &=& F_{i,21}= \frac {1}{P_{\text{D},i}(1-P_{\text{D},i})} \frac {\partial {P_{\text{D},i}}}{\partial r_i} \frac {\partial {P_{\text{D},i}}}{\partial P} \cos \psi_i,  \nonumber \\
F_{i,13} &=& F_{i,31}= \frac {1}{P_{\text{D},i}(1-P_{\text{D},i})} \frac {\partial {P_{\text{D},i}}}{\partial r_i} \frac {\partial {P_{\text{D},i}}}{\partial P} \sin \psi_i,  \nonumber \\
F_{i,23} &=& F_{i,32}= \frac{1}{P_{\text{D},i}(1-P_{\text{D},i})}\left(\frac {\partial {P_{\text{D},i}}}{\partial r_i} \right)^2    \sin \psi_i \cos \psi_i. \nonumber
\end{eqnarray}
\section{Derivation of $\mathbf{F}$} 
We have
\begin{eqnarray}
\mathbf{F}&=&-E_{\{[x_i~y_i]\}}E_{\{d_i\}}\left[ \frac {\partial^2  \ln p(\{d_i\}; \boldsymbol{\theta})}{\partial \boldsymbol{\theta}^2} \right] \nonumber \\
&=&-E_{\{[x_i~y_i]\}} \left[ \sum_{i} P_{\text{D},i} \frac {\partial^2 \ln P_{\text{D},i}}{\partial \boldsymbol{\theta}^2}+ P_{\text{ND},i} \frac {\partial^2 \ln P_{\text{ND},i}}{\partial \boldsymbol{\theta}^2}   \right] \nonumber \\
&=&-  \sum_{i} E_{\{[x_i~y_i]\}}\left[ P_{\text{D},i} \frac {\partial^2 \ln P_{\text{D},i}}{\partial \boldsymbol{\theta}^2}+ P_{\text{ND},i} \frac {\partial^2 \ln P_{\text{ND},i}}{\partial \boldsymbol{\theta}^2}   \right]. \nonumber
\end{eqnarray}
If we denote the region where sensors are distributed by $\mathcal{S}$ and area of that region by $A$, the expected number of sensors in $\mathcal{S}$ is $\rho A$ and the probability density function of a sensor being at location $[x,y]$ is $\frac{1}{A}$ if $[x,y]\in\mathcal{S}$ and zero otherwise. Therefore,
\begin{eqnarray}
\mathbf{F}&=&-\sum_{i=1}^{\lfloor	\rho A\rfloor }   \int_{\mathcal{S}} \frac{1}{A} \left( P_{\text{D}} \frac {\partial^2 \ln P_{\text{D}}}{\partial \boldsymbol{\theta}^2}+ P_{\text{ND}} \frac {\partial^2 \ln P_{\text{ND}}}{\partial \boldsymbol{\theta}^2} \right) ds \nonumber
\end{eqnarray}
Here, we have assumed that region is large but not infinite. Now, is we grow $\mathcal{S}\to\mathbb{R}^2$, we get
\begin{eqnarray}
\mathbf{F}&=& - \int_{-\infty}^{\infty} \int_{-\infty}^{\infty} \frac{\rho A}{A} \left( P_{\text{D}} \frac {\partial^2 \ln P_{\text{D}}}{\partial \boldsymbol{\theta}^2}+ P_{\text{ND}} \frac {\partial^2 \ln P_{\text{ND}}}{\partial \boldsymbol{\theta}^2} \right) dx dy \nonumber \\
&=&- \int_{-\infty}^{\infty} \int_{-\infty}^{\infty} \rho \left( P_{\text{D}} \frac {\partial^2 \ln P_{\text{D}}}{\partial \boldsymbol{\theta}^2}+ P_{\text{ND}} \frac {\partial^2 \ln P_{\text{ND}}}{\partial \boldsymbol{\theta}^2} \right) dx dy  \nonumber\\
&=& -\int_{0}^{\infty} \int_0^{2 \pi}  2 \rho  \pi r   \left( P_{\text{D}} \frac {\partial^2 \ln P_{\text{D}}}{\partial \boldsymbol{\theta}^2}+ P_{\text{ND}} \frac {\partial^2 \ln P_{\text{ND}}}{\partial \boldsymbol{\theta}^2} \right)  dr d\phi.   \nonumber
\end{eqnarray}
\section{Derivation of the average distance of the closest detector to the target}
To calculate the average distance of the closest detector to the target, first, assume that $N$ detectors are uniformly distributed in a disk of radius $R$. Thus, the node density is $\rho =\frac {\pi R^2}{N}$, and the density function of the distance of each detector from the target is
\begin{eqnarray}
f(r)&=&\frac {2r}{R^2}u(r)u(R-r),  \nonumber
\end{eqnarray}
where $u(.)$ is the unit step function. Thus, the pdf of order statistic $r_{\min}$, the distance of the closest detector to target, is \cite{David2003OrderStatistics}
\begin{eqnarray}
f_{r_{\min}}(r)&=& \frac {2 \pi \rho r} {1- \frac{r^2}{R^2}} \left(1- \frac{r^2}{R^2} \right)^{ \pi \rho R^2} \left[u(r)-u(r-R)\right],   \nonumber
\end{eqnarray}
which, as $R \rightarrow \infty$, converges to a Rayleigh distribution with parameter $\sigma_{r_{\min}}=\frac {1}{\sqrt {2\pi \rho}}$. Thus $E[r_{\min}]= \frac {1}{\sqrt{4\rho}}$.
\section{Detail of derivation of closed form approximate of $F_{22}$ }
We have
\begin{eqnarray}
\label{Eafinalaccurateinteg3}
F_{22}=F_{33}&\approx& \int_{\breve{r}}^{\infty} \int_0^{2 \pi} 2 \pi \rho r \frac {(\frac {\partial P_{\text{D}}}{\partial r})^2}{P_{\text{D}} (1-P_{\text{D}})} \cos^2\psi d \psi dr         \nonumber \\
\label{EaInt1overr}
&=&2 \pi^2 \rho \int_{\breve{r}}^{\infty}  \frac {(\frac {\partial P_{\text{D}}}{\partial r})^2}{P_{\text{D}} (1-P_{\text{D}})} rdr,\nonumber
\end{eqnarray}
which using change of variables
\begin{eqnarray}
x&=&\sqrt{\frac {T P}{\sigma^2 r^{\alpha}} },  \nonumber
\end{eqnarray}
and
\begin{eqnarray}
t&=&\sqrt{\frac {2\tau}{\sigma^2}}, \nonumber
\end{eqnarray}
and considering that
\begin{eqnarray}
\frac{\partial {P_{\text{D}}}}{\partial r}&=& \frac{-\alpha t x}{2r}  e^{-\frac {t^2+x^2}{2}} I_1(tx), \nonumber \\
\frac{dx}{x}&=&\frac{-\alpha}{2} \frac{dr}{r}, \nonumber
\end{eqnarray}
we get
\begin{eqnarray}
\label{I22Generalexac2}
F_{22}&\approx&\pi^2 \rho \alpha t^2 \int_0^{\breve{x}}   \frac {I_1^2\left(xt\right)e^{-x^2-t^2}}{Q(x,t) (1-Q(x,t))} xdx, \nonumber
\end{eqnarray}
where $\breve{x}=\sqrt{\frac {T P}{\sigma^2 {\breve{r}}^{\alpha}} }$. Using change of variable $y=xt $ and defining similar $A, B$, and $C$ variables to those defined for $F_{11}$ in the paper results in
\begin{eqnarray}
\label{J1approximation2}
F_{22}&\approx&C\pi^2 \rho \alpha \int_0^{\breve{y}}   yI_1^2\left(y\right)e^{-(Ay+B)^2} dy, \nonumber
\end{eqnarray}
where $\breve{y}=\breve{x}t$. We also approximate $I_1^2(y)$ with the first $m+1$ terms of its Taylor expansion around zero
\begin{eqnarray}
\label{TaylorExpansionI1sq}
I_1^2(y) & \approx & \sum_{k=0}^{m} {2k+2\choose k} \left[\frac{y^{k+1}}{2^{k+1}(k+1)!}\right]^2. \nonumber
\end{eqnarray}
Thus $F_{22}$ can be approximated as
\begin{eqnarray}
F_{22} &\approx& C\pi^2 \rho \alpha \sum_{k=0}^m  {2k+2\choose k} \frac{1}{[2^{k+1}(k+1)!]^2} \int_0^{\breve{y}}   y^{2k+3} e^{-(Ay+B)^2} dy. \nonumber
\end{eqnarray}
Now, change of variables $s=Ay+B$ and $\breve{s}=A\breve{y}+B$ yields
\begin{eqnarray}
F_{22}&\approx&\frac{C\pi^2 \rho \alpha}{A^2} \sum_{k=0}^m {2k+2\choose k}\frac{1}{[(2A)^{k+1}(k+1)!]^2}\int_B^{\breve{s}}(s-B)^{2k+3}e^{-s^2}ds \nonumber\\
& = & \frac{C\pi^2 \rho \alpha}{A^2} \sum_{k=0}^m {2k+2\choose k}\frac{1}{[(2A)^{k+1}(k+1)!]^2}\sum_{l=0}^{2k+3}{2k+3\choose l}(-B)^{l}\int_B^{\breve{s}} s^{2k+3-l}e^{-s^2}ds  \nonumber\\
& = & \frac{C\pi^2 \rho \alpha }{2A^2} \sum_{k=0}^m {2k+2\choose k}\frac{1}{[(2A)^{k+1}(k+1)!]^2}    \nonumber\\
& & \times \sum_{l=0}^{2k+3}{2k+3\choose l}(-B)^{l} \left[ \Gamma\left(k+2-\frac{l}{2},B^2\right)- \Gamma\left(k+2-\frac{l}{2},{\breve{s}}^2\right) \right].\nonumber
\end{eqnarray}

\end{document}